# Optical designs for the Maunakea Spectroscopic Explorer Telescope

Will Saunders[1] and Peter Gillingham,

Australian Astronomical Observatory, PO Box 915, North Ryde, NSW 1670, Australia

## ABSTRACT

Optical designs are presented for the Maunakea Spectroscopic Explorer (MSE) telescope. The adopted baseline design is a prime focus telescope with a segmented primary of 11.25m aperture, with speed f/1.93 and 1.52° field-of-view, optimized for wavelengths 360-1800nm. The Wide-Field Corrector (WFC) has five aspheric lenses, mostly of fused silica, with largest element 1.33m diameter and total glass mass 788kg. The Atmospheric Dispersion Corrector (ADC) is of the compensating lateral type, combining a motion of the entire WFC via the hexapod, with a restoring motion for a single lens. There is a modest amount of vignetting (average 5% over the hexagonal field); this greatly improves image quality, and allows the design to be effectively pupil-centric. The polychromatic image quality is $d_{80}<0.225''/0.445''$ at ZD 0/60˚ over more than 95% of the hexagonal field-of-view. The ADC action allows adjustment of the plate-scale with zenith distance, which is used to halve the image motions caused by differential refraction. A simple design is presented for achieving the required ADC lens shifts and tilts.

A two-mirror design was also undertaken for MSE, but was not selected. This is a 12.3m F/2.69 forward Cassegrain design, with a 2.75m diameter M2, and three silica lenses, of largest diameter 1.33m. The field-of-view is again 1.52°. The f/0.95 primary makes the design remarkably compact, being under 10m long. The ADC action involves a small motion of M2 (again via a hexapod), and shifts and tilts of a single lens. The design is effectively pupil-centric, with modest vignetting (5.9% average). The image quality is virtually identical to the prime focus design.

## 1. INTRODUCTION

The proposed Maunakea Spectroscopic Explorer (MSE) telescope [1] has an AΩ of $117m^2deg^2$, about twice that of the most ambitious current wide-field spectroscopic telescope projects ( DESI, PFS), and with a wider wavelength range too. Hence any prime focus WFC design for MSE can be expected to involve many large lenses, of multiple glass types, and with strong curvatures and asphericities. Fused silica is the strongly preferred lens material, because of its transparency, homogeneity and availability in large sizes. The Atmospheric Dispersion Corrector (ADC) is especially challenging for large WFCs, because it traditionally requires doublets and/or prisms of glasses other than fused silica. We have recently developed a Compensating Lateral ADC (CLADC) concept that allows an ADC action for almost any WFC, with no additional lenses [2,3,4]. The idea is that translating a lens laterally introduces a prismatic effect, which can correct for atmospheric dispersion, but also adds wavefront tilt and astigmatism; translating one or more other lenses in the system can correct for the tilt and astigmatism, leaving a pure ADC action. The concept works for both prime focus and two-mirror telescopes, and (at least in principle) has no impact on either throughput or monochromatic image quality.

Prime Focus (PF) and Forward Cassegrain (FC) CLADC designs were laid out for MSE in mid-2015. The PF CLADC design concept was adopted by the MSE project office in October 2015. The compactness of the FC design was not needed, while the PF design was cheaper and allowed a smaller etendue for equivalent survey speed (and hence cheaper spectrographs also); it also avoided the difficulty of a large aspheric convex secondary. It was successfully externally reviewed in February 2016. The design presented here is the current evolution of that PF design. The FC design is presented as an appendix.

## 2. MSE REQUIREMENTS

The relevant MSE requirements were as follows:

---

[1] will@aao.gov.au

- A segmented M1, with hexagonal segments of size 1.44m corner-to-corner (as for TMT) or less.
- Focal length 21.66m or greater (to allow a choice of positioner technologies).
- Total telescope length < 22m (to fit in the proposed calotte enclosure ).
- L1 no larger than 1.5m diameter, other lenses no larger than 1.2m, including mounting allowance.
- Wavelength range 370nm-1300nm as a requirement, and 360nm-1800nm as a goal.
- Required field diameter 1.52° (to give an area of 1.5deg$^2$ for the hexagonal field-of-view).
- $d_{80}$ < 0.35″/0.45″ at ZD 0/ZD 50° over 90% of the field radius, and no more than 20% worse at larger radii.
- Maximum zenith distance at least 60°, with 'graceful degradation in optical performance' beyond 50°.
- Vignetting < 8% (requirement, with 5% goal) at 90% radius, < 15% at full radius.
- Plate scale constant to ±3%.

We included the additional requirements:
- The aperture diameter of L1 was fixed to 1.25-1.3m, with sensitivity testing to size variations.
- The clearance in front of the focal surface was required to be ≥100mm (to allow for a future IFU upgrade).
- Non-silica lenses were restricted to the maximum blank size offered by Ohara with guaranteed homogeneity.

## 3. OVERVIEW OF PRIME FOCUS DESIGN

We started from a (reverse-engineered) version of the Subaru HSC design [5], which in turn derives from earlier Terebizh designs [6]. The HSC design has superb image quality, but also has 7 lenses, three glass types, vignetting over almost the entire field, and a flat but strongly non-telecentric focal surface. Both $\theta$-$\phi$ and tilting-spine positioner systems now allow curved focal surfaces [7, 8], so a flat focal surface is not a requirement for MSE. The sliding doublet ADC was removed, silica was substituted as far as possible, and the lens sizes were increased to meet the MSE vignetting specifications. The adopted design has just 5 lenses, with all the thickest and most powerful lenses of fused silica, and just two thin and weakly powered lenses of light flint. Various combinations of lens movements were tried for the CLADC action, and a solution was found that required just one mechanism in addition to the already planned hexapod.

WFCs for fiber use should in principal be pupil-centric, so as not to cause geometric FRD at fiber input. However, if there is vignetting at large field radii, this can be creatively used to clip light that would otherwise enter the fibers at the fastest speeds, allowing effectively pupil-centric designs (in the sense that a fiber and collimator speed which suffices at the center of the field suffices everywhere), even when the chief ray is not orthogonal to the focal surface. This idea has been independently proposed elsewhere [9].

In the course of this work, it was realised that the CLADC action readily allows small changes in plate scale. Differential refraction causes image motion over time for wide-field telescopes, with a shear pattern dependence on field position. By adjusting the plate-scale with zenith distance, the maximum image motion can be halved, without any additional cost or complexity. The CLADC action also introduces changes in the distortion pattern of the telescope, which are strongly dependent on the detailed CLADC design. The optimization minimized the combined image motion from both effects.

## 4. OVERALL LAYOUT

Figure 1 shows the layout of telescope and WFC. M1 consists of 60 segments, with overall diameter of 11.25m. M1 has focal length 18.770m, and is conic with $k = -1.113$. The overall length of the optics (M1 vertex to focal surface vertex) is 19.03m. The effective focal length is 21.66m. The focal surface is convex toward M1, with 10.24m RoC, and diameter 583.5mm. The global plate scale is 106.6μm/″ (focal surface diameter divided by field-of-view diameter), with ±2% variations. The final focal ratio is f/1.926 on-axis, which allows direct use with high Numerical Aperture (but still pure silica cored) fibers. The total effective telescope area (without allowance for vignetting by top-end or spiders) is 80.81m$^2$, for an effective diameter of 10.14m.

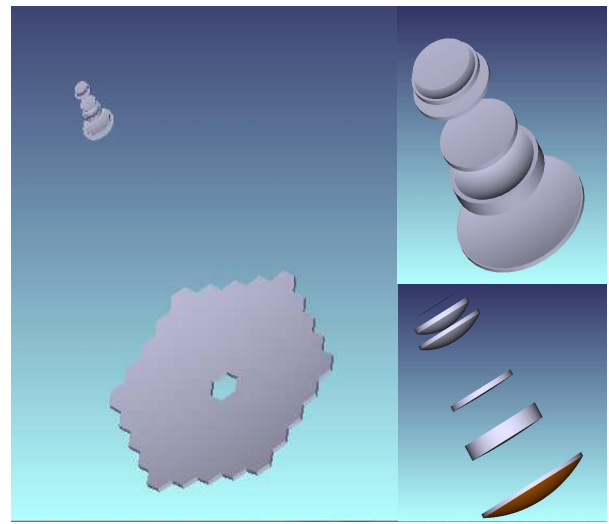

**Figure 1. Layout of the telescope and WFC**

## 5. WFC DETAILS

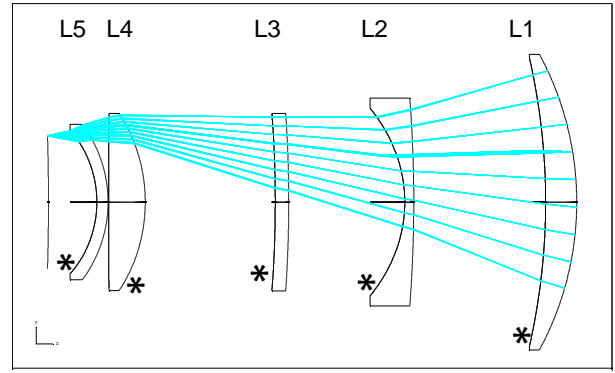

Figure 2 shows details of the WFC design. The three lenses with significant power (L1, L2, L4) are all fused silica, while L3 and L5 are of PBM2Y[2]. Each lens has an asphere, as shown (*). The total lens thickness is 325mm of silica and 110mm of PBM2Y. The total glass mass, including allowance for mounting, is 788kg. The total length of the WFC is 2.35m.

L1 is fused silica, with aperture diameter fixed at 1300mm. It is mildly aspheric (max aspheric slope 4mrad). Attempts to eliminate the asphere were unsuccessful.

L2 is fused silica, with 915mm aperture diameter. The concave surface is very strongly curved (RoC 601mm) and very strongly aspheric (max aspheric slope 49mrad, 2mm removed material), but it is almost a pure conic for easier testing.

**Figure 2. MSE WFC design. L3 and L5 are PBM2Y; all other lenses are fused silica. Surfaces marked with a * are aspheric.**

L3 is of PBM2Y, with 790mm aperture diameter, and max aspheric slope $7\times10^{-3}$. Efforts to dispense with this very weak lens were unsuccessful.

L4 is fused silica, with 789mm aperture diameter. The first face is strongly curved (RoC 706mm), and aspheric with maximum aspheric slope 33mrad. The closeness to focus allows a loose tolerance on the aspheric figure, which would be tested in reflection, through the lens.

L5 is of PBM2Y, with 683mm aperture diameter, a strong meniscus shape, and max aspheric slope 43mrad on the 2nd surface. The closeness to focus again allows a loose tolerance in the figure.

## 6. CLADC OPTICAL ACTION

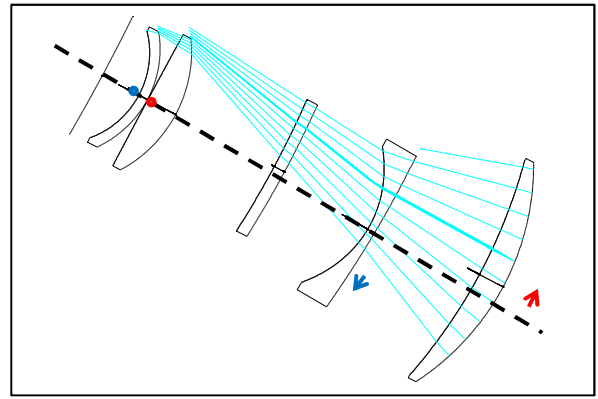

Figure 3 shows the Compensating Lateral ADC action at ZD 60°. The ADC action is achieved by **(a)** a shift and tilt of the hexapod, giving a rotation of the entire WFC about a horizontal axis just behind the focal plane; and **(b)** a compensating rotation of L2, giving a counter-rotation about almost the same axis. So L2 stays in almost the same place, while all the other components move. The required hexapod motion amounts to 7.2mm at L1, and the required motion of L2 is -6.5mm. The total resulting change to the telescope pointing is 8″.

The CLADC allows a deliberate change of plate-scale, to partially correct for distortion changes due to differential refraction (Section 8). This is via **(c)** an axial shift of L2 of 0.5mm towards M1, and **(d)** a compensating axial shift of the hexapod by 0.75mm away from M1 (to preserve overall focus).

**Figure 3. ADC action, at ZD 60°, exaggerated by a factor 10 for clarity. Dashed line shows the telescope axis. There is a rotation of the entire WFC about a horizontal axis behind the focal plane (red arrow and dot), with a compensating movement of L2 leaving it almost unmoved (blue arrow and dot).**

The tilt and lateral offset of L2 scale linearly with the differential refraction (i.e. roughly as $tan$(ZD), with a small correction for atmospheric curvature[3] [10]). The optimal axial offset of L2 has a quadratic dependence on $tan$(ZD). The L2 offset would be via an encoded actuator, while the precise hexapod motions would be generated automatically by the active optical system. A proposed mechanical arrangement to give the required motion of L2 is presented in Section 14.

---

[2] Schott LF5 or LLF1 could be used instead, but with some doubt about the homogeneity that can be obtained.

[3] The Zemax atmospheric model is incorrect for wide fields, because the same dispersion is applied all over the field. Instead, we directly included a simple model atmosphere in the telescope model. This consisted of a uniform atmospheric shell, of density, temperature and inner radius equal to the values at the telescope (0.609ATM, 10C, 7858km), and of thickness equal to the scale height (8km). This gave answers almost indistinguishable from actual measurement [10].

## 7. VIGNETTING

The principal vignetting losses come from the under-sizing of L1, which causes significant vignetting in the outer parts of the field, as shown in Figure 4(a).

Figure 4(b) shows the ZEMAX vignetting plot at zenith. The scale includes the effect of the non-circular M1. The loss due to WFC vignetting is 11.8% at the edge of the field, with area-weighted average (over the hexagonal area actually used for data-taking) of 5.0%. The ADC action causes a very small additional vignetting, at the top edge of the field.

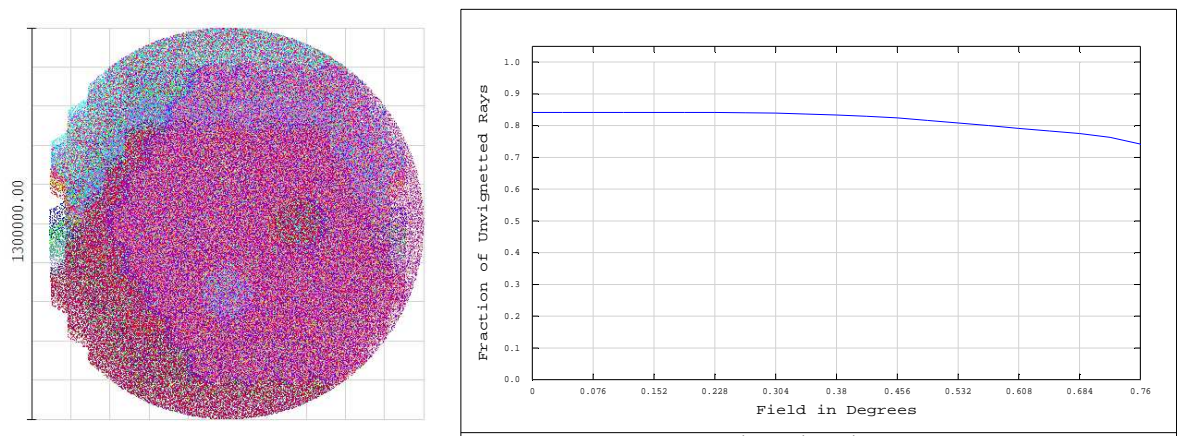

**Figure 4. (a) Beam profiles on L1 for each of the 12 field positions used. Dark red, pale blue and purple beams are all for targets at the edge of the field. (b) Overall vignetting, as calculated by ZEMAX, i.e with respect to a circular M1. Top-end and WFC vignetting are both included. There is progressive vignetting by L1 beyond ~0.4°, and by L3 beyond ~0.73°.**

## 8. DIFFERENTIAL REFRACTION AND DISTORTION CHANGES

In any wide-field telescope, differential refraction causes a variable ZD-dependent distortion, mostly a compression of the field in the elevation direction, i.e. a combination of monopole and quadrupole terms (Figure 5(a)). The CLADC action introduces additional changes to the distortion pattern of the telescope, of a typically smaller amplitude (depending on the detailed design) but more complex pattern. However, the ADC action also allows changes to the plate scale. This can be used to correct for the monopole term of the differential refraction, so the total change in distortion can be almost halved from that caused by differential refraction alone.

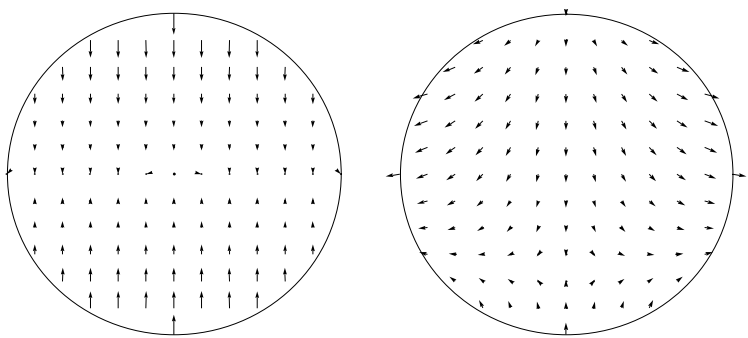

**Figure 5. Distortion change between ZD 0 and ZD 60°, (a) just due to differential refraction and (b) combined differential refraction and ADC action. The maximum distortion change is reduced from 153μm to 78 μm. The top of the field is nearest Zenith, and the vectors are exaggerated 250-fold.**

In practice, the optimal ADC action was determined by allowing L2 to shift axially along with translation and tilt, along with hexapod readjustment, and optimising the entire design with the resulting distortion changes minimised. Figure 5(b) shows the resulting combined effect of differential refraction and ADC action.

The largest residual image motion during a 40 minute frame at ZD<50° (and hence the largest open-loop fiber position correction ever normally required) is ~20μm.

# 9. IMAGE QUALITY

Figure 6 shows the polychromatic enclosed energy plots for (a) zenith and (b) ZD 50°. The lines show the requirements, in each case, both over the central 90% of the field radius (green), and at all radii (red). The design easily meets all four requirements, with the image quality over the whole field meeting the requirement within the central 90% field radius.

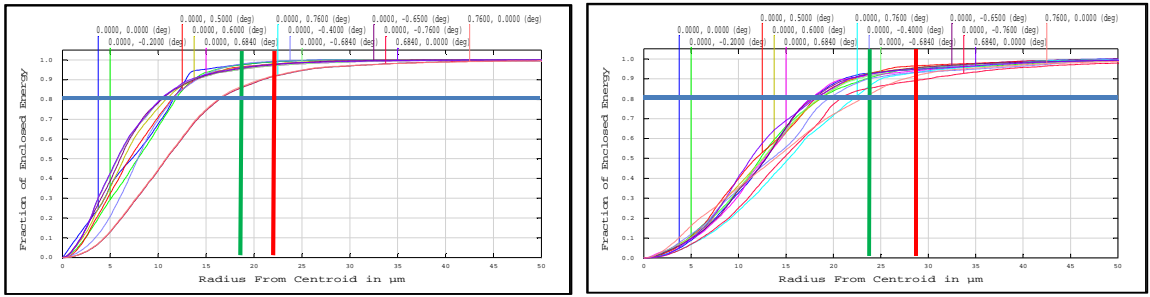

**Figure 6. Cumulative geometric enclosed energy plots, for (a) ZD 0 and (b) ZD 50°. The blue lines show the 80% level, the green lines show the requirement for the inner 90% of the field of 0.35″ and 0.45″ respectively, and the red lines show the requirements (20% greater) beyond this. The design meets all four requirements.**

Figure 7 shows the spot diagrams for wavelengths 360-1800nm, for (a) zenith and (b) ZD 50°. The circles are 100μm (0.95″). diameter. It can be seen that the monochromatic image quality remains good at ZD 50°; the polychromatic image quality degradation is mostly caused by secondary spectrum from the ADC action. Figure 8 shows the polychromatic rms image radius (in mm) over the field, at zenith and ZD 50°, showing the asymmetry caused by the ADC action.

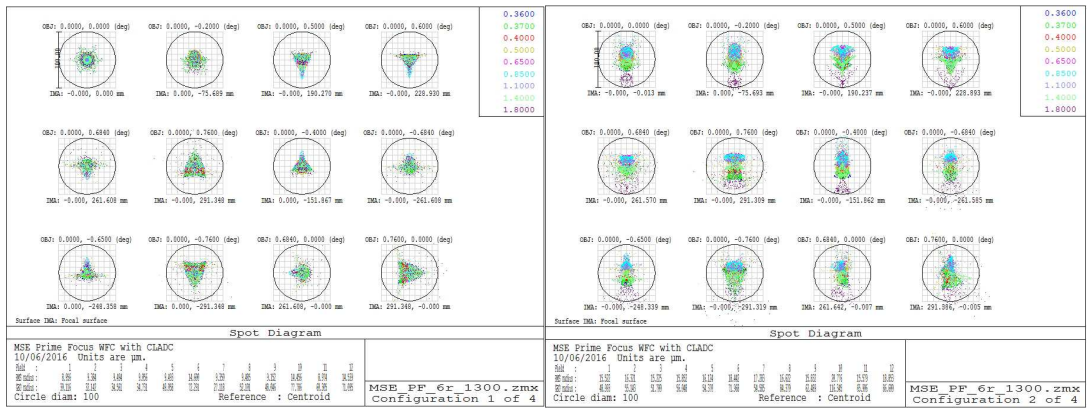

**Figure 7. Spot diagrams for (a) ZD 0, (b) ZD 50°, for 360, 370, 400, 500, 650, 850, 1100, 1400, 1800nm. The circles are 100μm (0.95″) diameter. The three worse fields (#6, 10, 12) are all at the full field radius of 0.76°.**

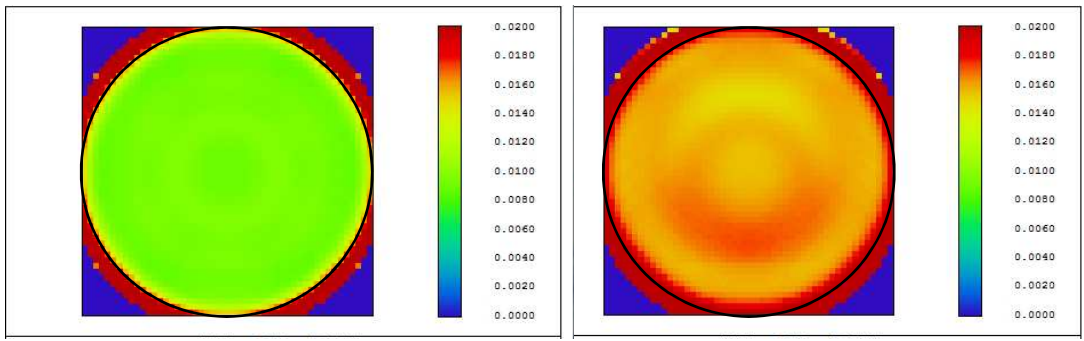

**Figure 8. Rms image quality maps for (a) ZD 0 and (b) ZD 50°. The black circle indicates the 1.52° field. Typical spot rms radii are 15μm (0.14″) at ZD 0, and 20μm (0.19″) at ZD 50°, but showing the asymmetry caused by the ADC action.**

## 10. TELECENTRICITY

The design is not formally pupil-centric, with the chief rays being up to 2.1° away from the normal to the focal surface at ZD 0, and up to 2.5° away at ZD 60°. Normally this would cause geometric FRD, with some light entering the fibers at greater angles than required for use on-axis at Zenith. However, the vignetting at L1 very efficiently cuts out these fastest rays. Figure 9 shows the far-field illumination of the fibers, for (a) ZD 0, and (b) ZD 50°. The circle is the minimum speed required to enclose all rays for the center of the field at zenith. The same circle encloses essentially all the light at all field positions and zenith distances. This makes the design effectively pupil-centric, i.e. the fiber and collimator acceptance speeds required for use on-axis at zenith will also suffice at all other field positions and zenith distances.

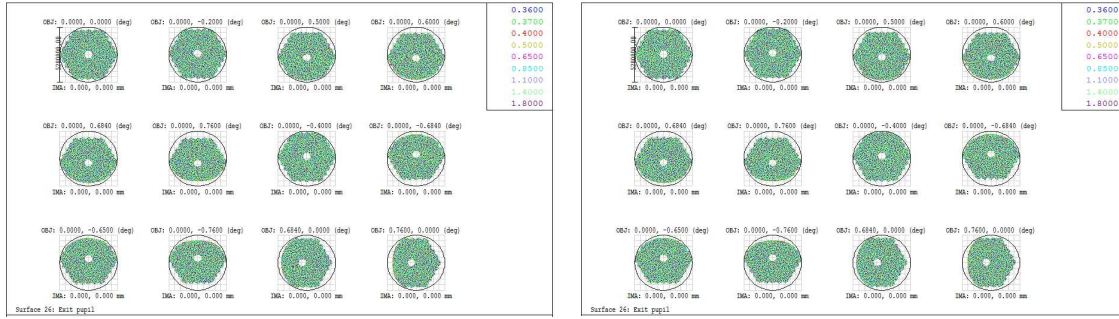

**Figure 9. Far-field illumination of the fibers, for the 12 fields used in the optimization, at (a) ZD 0 and (b) ZD 50°. The circle represents the telescope speed on-axis at Zenith (F/1.926); it also suffices at all field positions and zenith distances.**

## 11. COATINGS AND THROUGHPUT

The WFC anti-reflection (AR) coatings are a major issue for MSE, because of the huge (5-fold!) wavelength range, and because of the strong curvature of some of the lenses. None of the multilayer hard coatings that were offered had good performance. However, an excellent alternative is offered in the form of spin-coated solgel, especially over $MgF_2$, with binding, cleaning and hydrophobic layers as needed [11]. Varying the design thickness of the two coats allows tuning of the two wavelengths with least losses, a strong benefit for MSE which is S/N limited in both the blue (for stars) and far-red/NIR (for faint galaxies). The thickness of solgel is determined by surface tension, and so should not vary greatly with surface slope. The $MgF_2$ thickness does depend on slope, but masking can be used to minimize the variations. L1 is somewhat larger than the largest optic solgel coated to date (the Keck ADC at ~1.2m); the other lenses are all smaller.

The broadband AR performance of solgel+$MgF_2$ is excellent, and the glass transparency is excellent with only 110mm of glass other than fused silica. Figure 10 shows the theoretical throughput of the WFC, for center and edge of the field, and the average over the hexagonal FoV. Included are vignetting, glass transmission, and coatings. The theoretical average throughput is better than 90% at ~500nm and ~1000nm, and is > 80% for 400nm-1500nm. Not included are the effects of binding, cleaning and hydrophobic layers, or any coating thickness errors or variations.

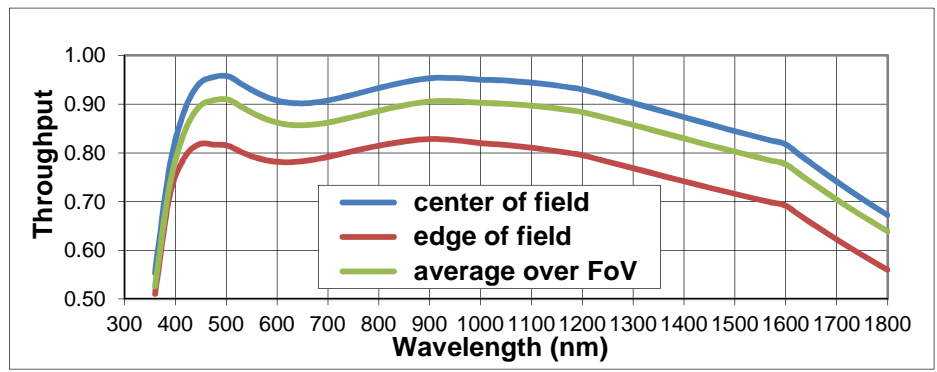

**Figure 10. Theoretical WFC throughput, including lens vignetting, glass transparency, and AR coatings**

## 12. VARIANTS

Earlier versions of this design had L1 closer to focus by about 150mm, which, for fixed L1 aperture, gives less vignetting (by ~2% on average). However, moving L1 away from focus improves the image quality dramatically, because (a) the image quality improves even at fixed vignetting (i.e. with larger L1), and (b) it increases further when the vignetting is increased (i.e. restoring L1 to its original size). The WFC image quality is one of the major components of the Facility Image Quality (FIQ), and the superb natural seeing demands the best possible FIQ. A better FIQ also allows smaller fibers, reducing the (very large) spectrograph costs.

The imaging and vignetting specs could be just met with a somewhat smaller L1, of ~1200mm aperture, but at the cost of ~30% worse image quality, as well as even stronger asphericities and radii of curvature. An L1 of 1400mm aperture would have ~10% better image quality and slightly gentler asphericities and RoCs.

Bernard Delabre has provided strikingly simple variants of this design, using the same CLADC principal, but with only 4 or even 3 lenses, and of weaker asphericities, and with much better back-focus. Simpler designs have a hidden benefit, in that the image quality is likely to be closer to the Zemax model than for a more challenging design. However, to date, none of the simplified designs meet the imaging specification. They also have very powerful lenses not of fused silica, raising issues of availability and homogeneity.

## 13. TOLERANCING

With its ambitious performance specifications, particularly as regards aperture and field diameters and wavelength range, this WFC was always bound to be challenging to manufacture. A very preliminary tolerancing analysis (mainly using the Zemax provisions for this) was performed on a late 2015 version of the design. Since then there have been several detailed changes but none that should significantly alter the conclusions reached then.

Tests were made in which tolerances on all surface curvatures, glass thicknesses, surface and element decentres were iteratively adjusted so almost all made roughly equal contributions to image degradation. Focal surface tip and tilt compensation was allowed. Compensations (mainly for curvature errors) were allowed by adjusting all air thicknesses, and also the RoC of surface L4/2 (which is in itself very tolerant to RoC error). Glass inhomogeneities equivalent to refractive index ranges of $\pm 2 \times 10^{-6}$ for the first two elements (equivalent to Schott's and Ohara's second highest qualities) and $\pm 10 \times 10^{-6}$ for the final three (twice as great as for the third highest qualities) were simulated with Zernike polynomials on one surface of each element.

Figure 11 shows the effect on the rms radius map for the worst Monte Carlo result out of ten. It is striking that the worst degradation is at the edges of the field, where (because of the hexagonal field-of-view) it has least effect. The overall degradation in image size, over the inner 90% of the field radius, is ~11% on average and ~16% in the worst case, representing (assuming quadrature) additional tolerancing errors of 6-8μm in rms radius, or $d_{80}$ ~ 0.07″-0.10″.

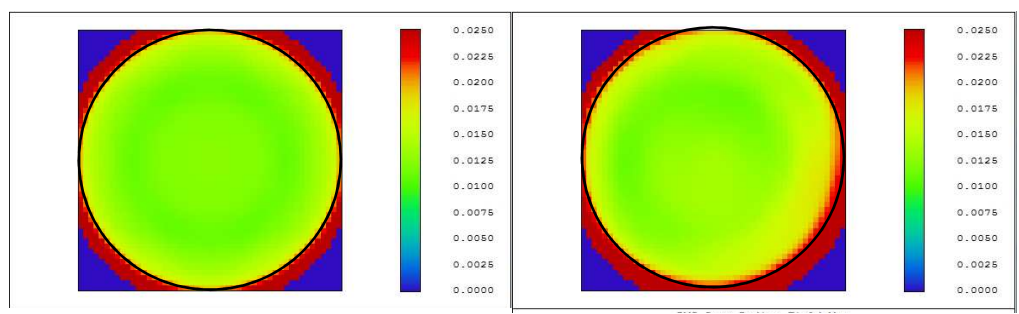

**Figure 11. RMS radii maps for (a) the nominal error-free design and (b) the worst of 10 Monte Carlo trials.**

The nominal tolerances on curvature remain very stringent, $\pm 10^{-7}$/mm for L1/L2/L3, and twice this for L4/L5. The resulting sagitta tolerance is e.g. ~20μm on L1. They could be significantly relaxed if re-optimisation of other curvatures was allowed after measurement of the achieved curvature on the most critical surfaces during the manufacturing process. Similarly, the tolerances on eccentricity could be relaxed if compensating adjustments during assembly, based on accurate measurements, were assumed.

Errors in the asphericities (beyond spherical aberration and astigmatism), were not included in the analysis described so far. Insofar as the greatest polishing errors on the aspheric surfaces are likely to be axisymmetric ripples in the radial profile, the aspheric errors were simulated by applying the Zemax surface type "extended cubic spline" on one surface of a dummy thin element (otherwise a plane parallel plate) next to L2. Since the slope departure of the aspheric surface from the best fitting sphere is bound to be greater at large radii, the radial error profile was simulated as a quasi-sinusoid with pitch diminishing linearly from a maximum at the centre. The plot of rms image radius vs radius in the field was slightly degraded when the simulated ripples had maximum slopes equivalent to 11μrad. It was proposed that this slope be adopted as the nominal tolerance for the aspheric surface on L2. For the surface on L1, allowing for its greater distance from focus, an equivalent tolerance would be ± 8μrad. For L3 it could be ± 16μrad and for L4 and L5 ± 50μrad. These are extremely tight tolerances and they will have to be re-visited thoroughly in the more comprehensive tolerancing study that is to begin shortly.

## 14. PROPOSED ADC MECHANISM

We have found a compact, robust, and versatile mechanism for controlling the required L2 offsets for the CLADC. The support against gravity is provided by three 25 mm diameter cam followers with needle roller bearings, as shown in Figure 12. These run in slots allowing as much travel as needed for lateral offsetting of the lens (by up to 6.6 mm at ZD 60º).

Two similar mechanisms would control the position angle of the lens, and a linear actuator with integral encoding would drive the lens offset, giving the required six constraints on the lens position. The layout is shown in Figure 13.

As modelled here, the slots **A**, **B**, and **C** have circular profiles constraining the lens to rotate around a fixed horizontal axis as it is offset laterally, as suggested in Figure 14. By changing the slopes and curvatures of the slot profiles, an axial shift can be introduced at the same time. This is desirable to partly compensate for the change of atmospheric distortion with changing elevation during an observation (Section 8).

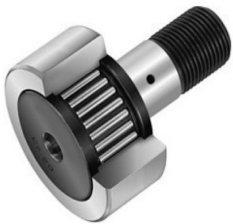
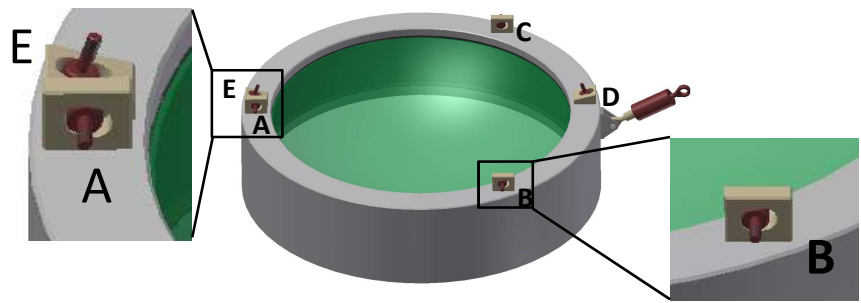

**Figure 13. Commercial cam follower**

**Figure 12. Center: General view of lens L2 in cell with support against gravity provided by three cam followers (A, B, and C) in short tracks and control in position angle by two similar arrangements (D and E). The cam follower shafts are assumed to be rigidly located in the main WFC structure. Left and Right: Close up views of the roller supports at A and B,**

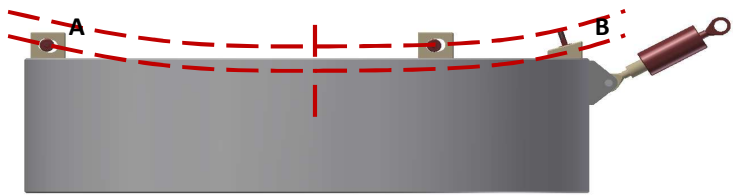

**Figure 14. Elevation view (with telescope at the zenith). The radial guide and roller at E are made invisible so as to show the arrangement at A more clearly. The two dashed arcs have their centres at the desired axis of rotation of the lens element for the ADC action. The angle of the ramp at A is roughly twice that of the ramp at B. By adjusting the profiles of the guides, an axial motion can be added to the rotation.**

# APPENDIX. CASSEGRAIN DESIGN

The first design we laid out for MSE was a somewhat larger (12.3m) two-mirror design, inspired by the VISTA telescope [12]. We have revisited this design, in the light of experience gained from the PF design. It also includes a CLADC, as also proposed (but not adopted) for 4MOST [3]. Because a two-mirror design gives good aberration correction even without a WFC, the WFC can have fewer lenses and glass types. We found a very satisfactory all-silica 3-lens design with an asphere on each lens. To get a focal surface physically small enough for wide-field correction with available lens sizes, drove the design to a very fast speed, f/2.69, with a primary of f/0.95, and a forward-Cassegrain focus, in front of M1.

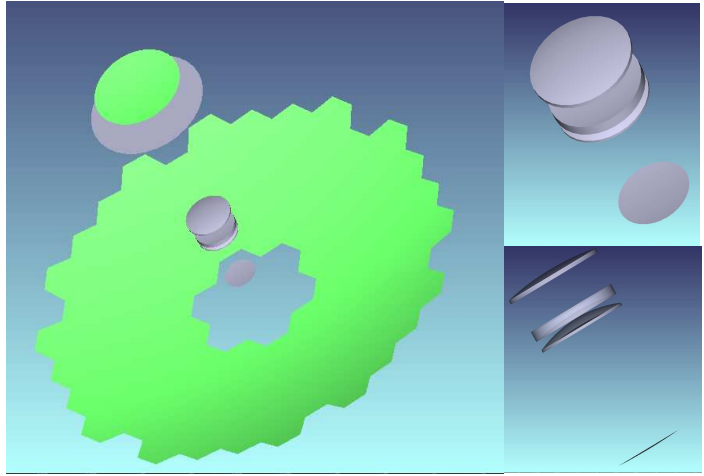

**Figure A1. Layout of 2-mirror telescope. M1 is segmented, ~f/1, with diameter 12.3m. M2 has working aperture 2.75m. The effective focal length is 36.565 m, and focal ratio f/2.97. The focal surface is ~1.4 m ahead of the M1 vertex, has diameter 896mm, and radius of curvature 7.5m.**

Like any 2-mirror telescope, Cassegrain and top-end baffles are required to prevent sky light reaching the focal surface directly (Figure A2(a)). The Cassegrain baffle costs no light, and serves to protect the primary when working on the instrument or WFC. The top-end baffle vignettes M1 directly, and the larger M2 is, the larger the top-end baffle required. Hence the best throughput is achieved for a slightly undersized M2 (Figure A2(b)), and this also reduces the etendue and consequent spectrograph costs. The adopted M2 diameter was 2.75m, and 3.6m for the baffle. The resulting loss (compared with the segmented primary) is 4.5% on-axis and 13% at the edge of the field, with average 5.9% over the hexagonal field-of-view. M1 and M2 are both aconic, close to $k = -1.043$ and $k = -8$ respectively. The focal surface has diameter 887mm, RoC -7.5m (concave as seen from M2), and the focal length is 33.34m, for an average plate scale of 162μm/″.

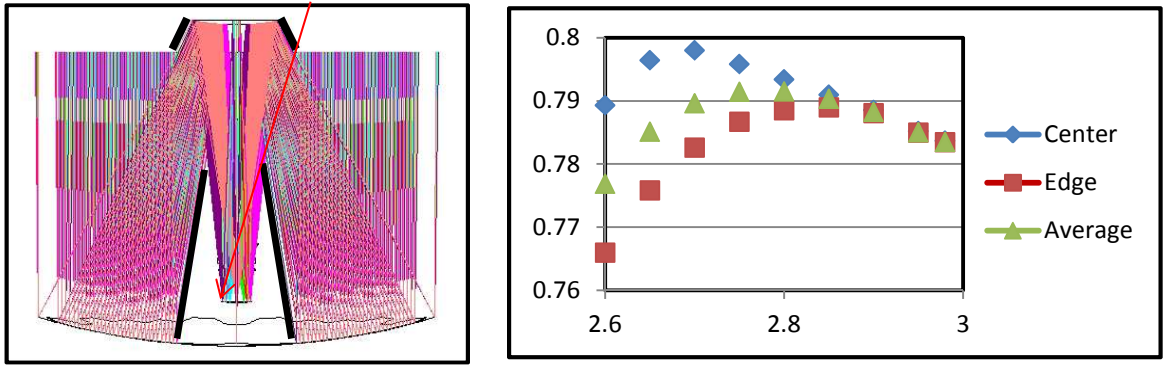

**Figure A2. (a) M2 baffles (thick black lines) to prevent direct sky light (red arrow) reaching the focal surface. (b) Overall throughput (compared with circular primary) as a function of M2 size, combining the effect of M2 overfilling and obstruction by the baffle.**

The WFC is shown in Figure A3. All three elements are of fused silica, with diameters, including 15mm radius mounting allowance, of 1330mm, 1201mm, 1193mm for L1/L2/L3. The total lens mass is 602kg. Each lens has a concave asphere, of maximum aspheric slope 20mrad, 28mrad, 1.6mrad for L1/L2/L3.

The ADC action consists of moving L3 laterally (against gravity), by up to 29mm, combined with a tilt of 35′ (in the sense of following the curve of the lens). There is also an axial motion of up to ~1mm, to compensate for differential refraction (see below). There is also a small (<3mm) movement of the M2 hexapod, which would be generated from a look up table generated, itself by the active optics system, and a large (but deterministic) pointing offset, of up to 3′. The motion of L3 causes additional vignetting of up to 2%, affecting just a 3′ wide crescent at the lowest edge of the field. The effect on the overall average throughput is negligible.

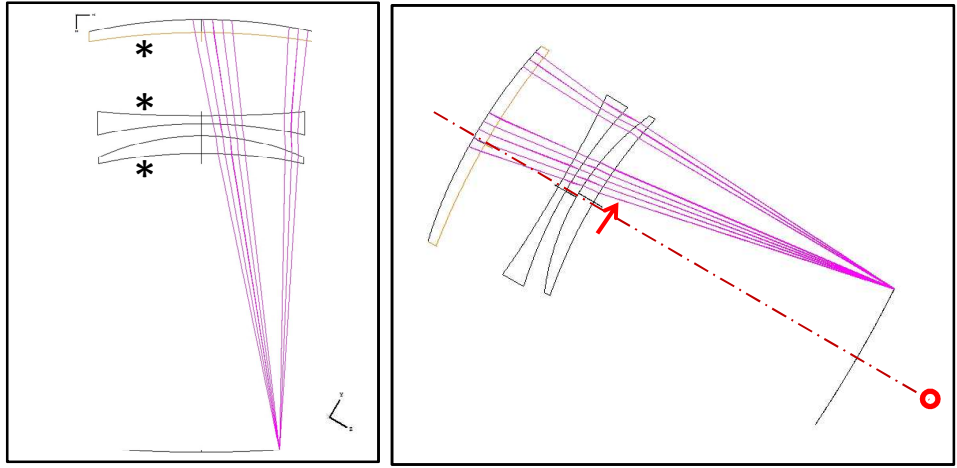

**Figure A3. Details of WFC (a) at zenith and (b) at ZD 60°. All three elements are of fused silica. Aspheres are marked as *. The maximum aspheric slope is 20mrad, 28mrad, 1.6mrad on Ll/L2/L3. At ZD 60°, L3 is rotated about the axis identified by the red circle clockwise through 35′, making the displacement of its vertex ~29mm. It also has a ~1mm axial motion to compensate for differential refraction.**

The image quality is as least as good as the Prime Focus design. Figure A4 shows the encircled energy distribution; at zenith, the image quality is $d_{80} < 0.21″$ over the 90% of the field radius.

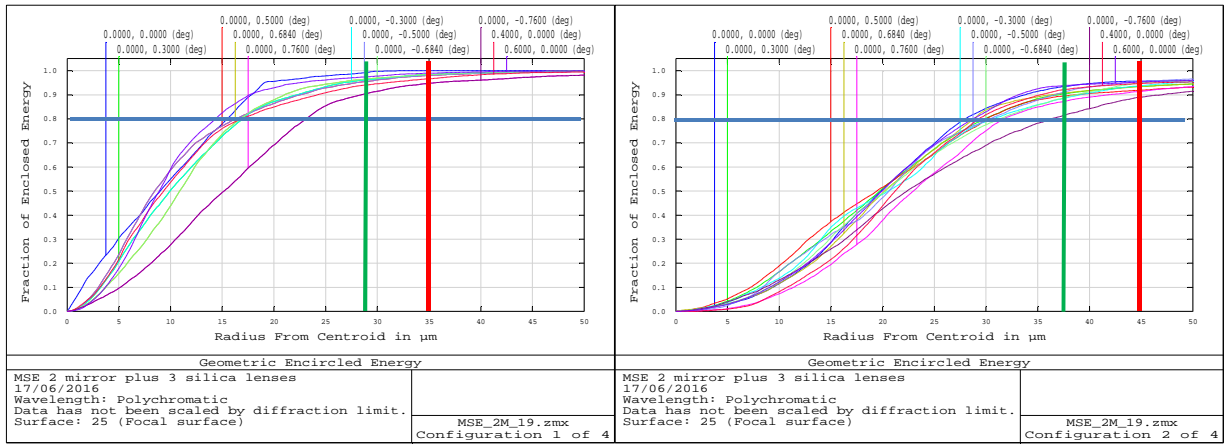

**Figure A4. Cumulative geometric enclosed energy plots, for (a) ZD 0 and (b) ZD 50°. The blue lines show the 80% level, the green lines show the requirement for the inner 90% of the field of 0.35″ and 0.45″ respectively, and the red lines show the requirements (20% greater) beyond this. The design meets all four requirements.**

The Figure A5 shows the spot diagrams, and A6 shows the rms image quality maps, all for ZD 0 and ZD 50°. The monochromatic image quality degrades at the edge of the field for large ZD's; this could be avoided by also moving L1, by a few mm, as part of the ADC action.

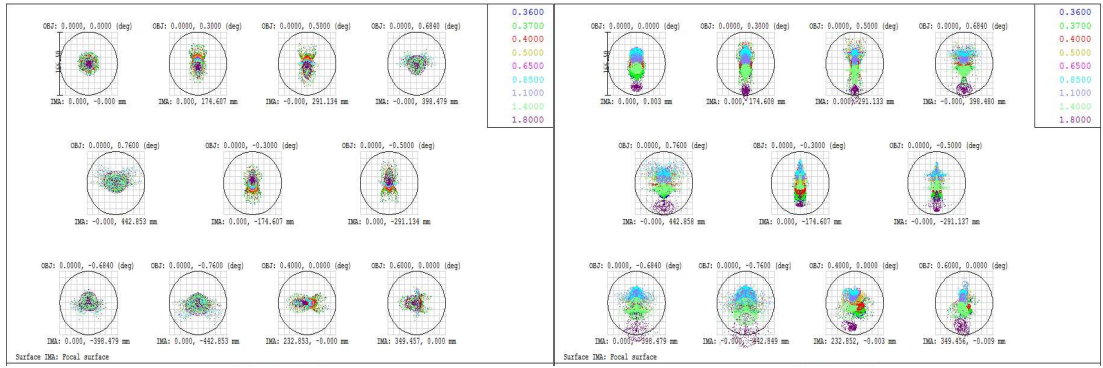

**Figure A5. Spot diagrams for (a) ZD 0, (b) ZD 50°, for 360, 370, 400, 500, 650, 850, 1100, 1400, 1800nm. The circles are 1″ diameter. The two worse fields (#5, 9) are at the full field radius of 0.76°.**

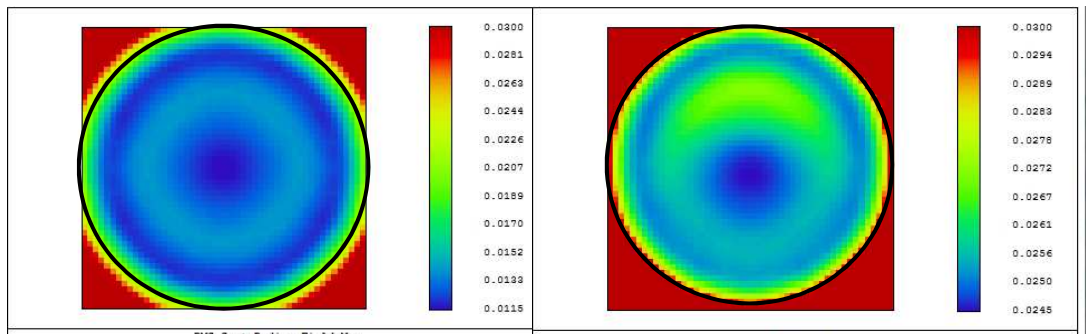

**Figure A6. Polychromatic rms image radius maps. (a) ZD 0, (b) ZD 50°. The highest value on the colour bar is 30 um (0.18″). The black circle shows the limit of the 1.52° diameter field.**

The vignetting makes the design effectively pupil-centric, much like the PF design. Figure A7 shows the far-field illumination of the fibers for ZD 0 and ZD 50°. Again, the fiber and collimator acceptance speed for on-axis use at zenith suffices everywhere.

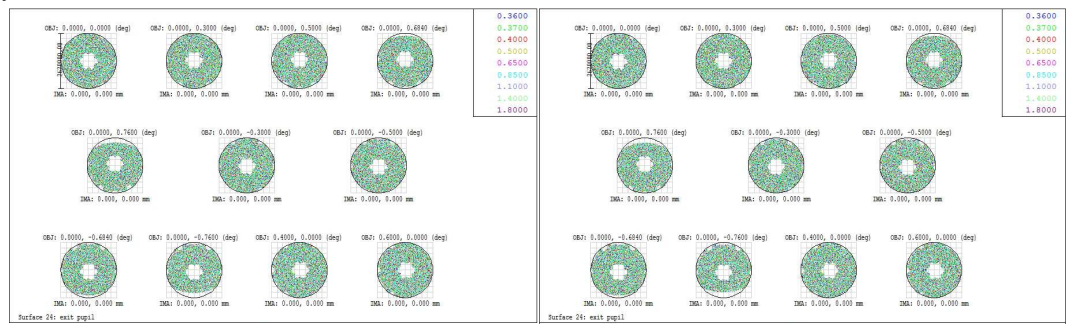

**Figure A7. Far-field illumination of the fibers, for the 12 fields used in the optimization, (a) ZD 0 and (b) ZD 50°. The circle represents the speed just accepting all the light on-axis at Zenith (F/2.69). The same acceptance speed accepts essentially all the light at all field positions and zenith distances.**

The ADC action in itself introduces small changes in the distortion pattern, <15μm between ZD 0 and ZD 60°. But, as part of the ADC action, L3 can be moved axially to change the plate scale, with negligible change to image quality. Hence image motion due to differential refraction can be halved, just as for the PF design.

The Cassegrain design presented here has the same field of view as the Prime Focus design, and even better image quality. A significant disadvantage is that it requires a larger primary and hence larger etendue for a given survey speed (because of the extra mirror and obstruction by the top-end baffle and M2), increasing the already large spectrograph costs; also the 2.75m diameter M2 is a significant cost and risk. However, the design is so compact (being shorter than it is wide), that the enclosure can be very small, a large cost saving for a new-build telescope. Also, an even larger diameter M1 could be considered. The limit comes from fused silica blank availability, at about 1.55m diameter. This would allow a ~15m primary of the same design, or even larger if the FoV is reduced.

## ACKNOWLEDGEMENTS

We are indebted to many people for the evolution of this design. Specifically, Paulo Spano, Damien Jones, Peter Doel, Andrew Rakich, John Pazder, Drew Phillips and Bernard Delabre have all given invaluable advice, along with MSE staff members Rick Murowinski, Kei Szeto, Shan Mignon, Derrick Salmon, Steve Bauman, Nicolas Flagey and Alan McConnachie.